\documentclass[aps,prb,onecolumn,superscriptaddress,floatfix,showpacs,amsmath,amssymb,nofootinbib,longbibliography]{revtex4-2}

\usepackage{graphicx}
\usepackage{dcolumn}
\usepackage{bm}
\usepackage{color}
\usepackage{tabularx}
\usepackage{csquotes}
\usepackage[normalem]{ulem}

\newcommand{\ncto}{Na$_2$Co$_2$TeO$_6$}

\newcommand{\tc}{$T_{\rm cr}$}
\newcommand{\tn}{$T_{\rm N}$}

\newcommand{\ac}{$\alpha_{\rm c}$}

\newcommand{\bc}{$B_\mathrm{C}$}

\begin{document}


\title{Competing Interactions and the Effects of Uniaxial Out-of-plane Perturbations in the Honeycomb Antiferromagnet Na$_2$Co$_2$TeO$_6$}

\author{J.~Arneth}\email{jan.arneth@kip.uni-heidelberg.de}
\affiliation{Kirchhoff Institute for Physics, Heidelberg University, INF 227, D-69120 Heidelberg, Germany}
\author{R. Kalaivanan}\affiliation{Institute of Physics, Academia Sinica, Taipei 11529, Taiwan}
\author{R. Sankar}\affiliation{Institute of Physics, Academia Sinica, Taipei 11529, Taiwan}
\author{K.-Y.~Choi}
\affiliation{Department of Physics, Sungkyunkwan University, Suwon 16419, Republic of Korea}
\author{R.~Klingeler}\email{klingeler@kip.uni-heidelberg.de}
\affiliation{Kirchhoff Institute for Physics, Heidelberg University, INF 227, D-69120 Heidelberg, Germany}

\date{\today}

\begin{abstract}
Despite exhibiting magnetic long-range order below $T_\mathrm{N} = 26.7$~K, the honeycomb cobaltate \ncto\ is predicted to enter a Kitaev spin liquid state when subjected to small external perturbations. While most of the reported literature investigates the effects of magnetic fields applied parallel to the honeycomb layers, we present high-resolution capacitance dilatometry studies for fields perpendicular to the Co-planes up to $15$~T. Grüneisen analysis reveals the effect of uniaxial out-of-plane strain and shows that antiferromagnetic order in \ncto\ is stabilized at a rate of $\partial T_\mathrm{N}/\partial p_\mathrm{c} = 0.28(5)$~K/GPa. Further, failure of the Grüneisen scaling at low temperatures around $T_\mathrm{cr} \simeq 7.5$~K demonstrates the presence of competing energy scales. In contrast to an only weak field dependence of the anomaly at \tn , a broad hump at \tc ($B=0$~T) evolves into a sharp peak at high fields applied $B \parallel c$. Our magnetostriction data show that a kink in the magnetisation at $B_\mathrm{C} \simeq 4.6$~T is accompanied by an inflection point in the field-induced length changes, which is likely related to weak unequal spin canting. All observed phenomena leave their signatures in the magnetoelastic phase diagram as constructed by our experimental results.
\end{abstract}
\maketitle

\section{Introduction}

Unlike in the prime example of frustrated magnetism, the triangular antiferromagnet, the exactly solvable Kitaev model introduces magnetic frustration not by the geometry of the underlying lattice but via strongly anisotropic bond-dependent magnetic interactions that can not be simultaneously satisfied for a honeycomb network of $S = 1/2$~\cite{kitaev2006}. At low temperatures, the resulting magnetic ground state of this model system is given by a quantum spin liquid (QSL) phase, which features fractionalized elementary excitations. Hence, it did not take long until the first suggestions of extremely fault-tolerant quantum computers based on materials realizing the Kitaev model were put forward~\cite{kitaev2003,duan2003,micheli2006}. However, so far, the search for actual compounds hosting a Kitaev quantum spin liquid ground state continues. After thorough investigation of the initial 'first wave' candidates, which were mainly based on $4d$ and $5d$ late transition metal ions with strong spin-orbit coupling~\cite{jackeli2009,singh2010,singh2013,nasu2016,takagi2019}, two independent theoretical studies predicted that sufficiently strong Kitaev interactions might emerge also in high-spin Co$^{2+}$ honeycomb lattice systems~\cite{liu2018_ncto,sano2018,liu2020}.

Among these 'second wave' compounds, \ncto\ has attracted the most attention by far~\cite{viciu2007,lefrancois2016,xiao2019,lin2021,hong2024}. Although antiferromagnetic long-range order emerges at $T_\mathrm{N} \simeq 26.7$~K, thermodynamic and spectroscopic studies suggest that \ncto\ can be described by an extended Heisenberg-Kitaev model with coupling parameters reasonably close to hosting a QSL ground state~\cite{songvilay2020,kim2021,samarakoon2021,sanders2022,krüger2023}. Such an assignment, thus, prompts the idea of using external perturbations, including magnetic and electric fields or pressure, to tune the balance of the magnetic interactions in favour of the Kitaev exchange. The bulk of the recent literature focuses dominantly on the influence of magnetic fields applied perpendicular to the Co-Co bonds within the honeycomb planes since the system undergoes a first-order phase transition into a putative spin disordered state at $B \parallel a^* \simeq 6$~T~\cite{lin2021,xiang2023,hong2024}. We have recently shown that this transition exhibits signatures of a quantum critical endpoint~\cite{arneth2025}. Yet, the exact magnetic state in the high-field phase of \ncto\ remains unclear.

In this work, we explore the effects of two, only scarcely investigated perturbations, namely \textit{uniaxial} pressure and magnetic fields applied perpendicular to the Co-layers~\cite{xiao2021,zhang2024,zhou2024_arxiv}. We employ high-resolution capacitance dilatometry which has been proven as a valuable method to elucidate and quantify spin-lattice coupling in layered correlated electron systems~\cite{he2018,gass2020,spachmann2022,arneth2022,spachmann2023}. Using Grüneisen analysis, we find competing energy scales at low temperatures and determine the uniaxial pressure coefficient of \tn\ from the experimental data at zero magnetic field. Magnetostriction and thermal expansion measurements for fields $B \parallel c$ up to $15$~T enable us to elucidate the role of magnetoelastic coupling in the thermodynamic properties of \ncto . We also present the magnetoelastic phase diagram, which corroborates the findings of our analysis.

\section{Experimental Methods}

All experiments have been performed on a high-quality 0.145~mm thin ($L_\mathrm{c}(300\,\mathrm{K},0\,\mathrm{T}) = 0.145$~mm) single crystal of \ncto\ grown as described in Ref.~\cite{lee2021}. The possible formation of an antiferromagnetic polymorph impurity phase~\cite{dufault2023} is excluded by our magnetisation data (see Fig.~\ref{fig:lowBChi_supp} in the Appendix). Out-of-plane relative length changes $dL_\mathrm{c}/L_\mathrm{c}$, i.e., along the crystallographic $c$-direction, were studied by means of a three-terminal high-resolution capacitance dilatometer (Kuechler Innovative Measurement Technology) in a home-built set-up placed inside a Variable Temperature Insert of an Oxford magnet system~\cite{Kuechler2012, Werner2017}. Magnetic fields were applied in the direction of the measured length changes, i.e.~$B \parallel c$. From the relative length changes the linear thermal expansion coefficient $\alpha_{\rm c} = 1/L_{\rm c}\times(\partial L_{\rm c}/\partial T)$ and the longitudinal magnetostriction coefficient $\lambda_{\rm c} = 1/L_{\rm c}\times(\partial L_{\rm c}/\partial B)$ were derived. Thermal expansion and magnetostriction measurements were performed at sweep rates of $0.3~$K/min and $0.3~$T/min, respectively. The magnetisation was studied using the vibrating sample magnetometer (VSM) option of a Physical Properties Measurement System (PPMS-14, Quantum Design).

\section{Experimental Results}

\begin{figure}[]
    \centering
    \includegraphics[width = 0.5\columnwidth]{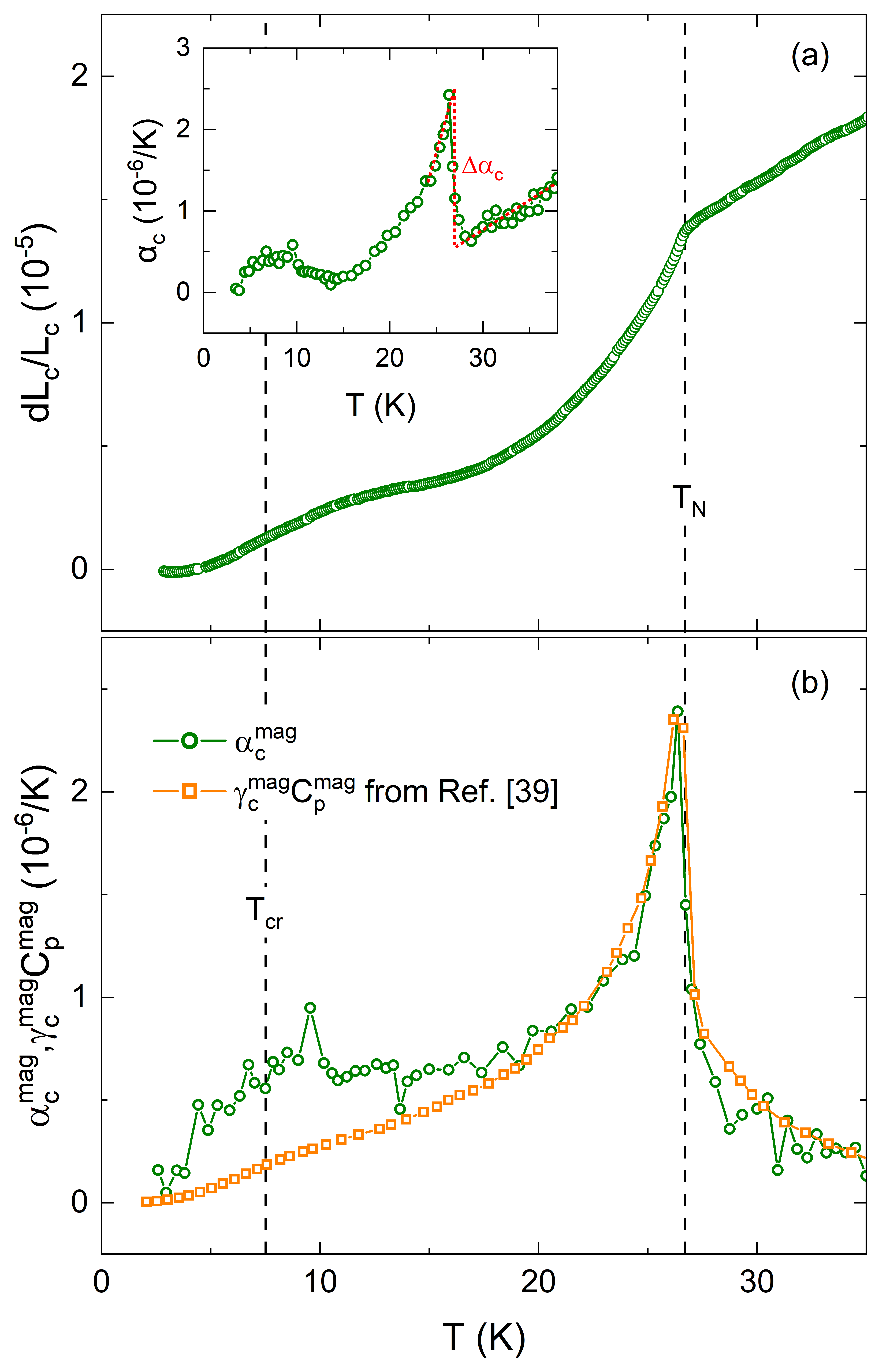}
    \caption{(a) Out-of-plane thermal expansion $dL_\mathrm{c}/L_\mathrm{c}$ and corresponding thermal expansion coefficient $\alpha_\mathrm{c}$ (inset) at $B = 0$~T. Dotted red lines in the inset illustrate the determination of the jump $\Delta\alpha_\mathrm{c}$, at \tn , employing an area-conserving construction. (b) Magnetic contribution to the thermal expansion coefficient and to the specific heat capacity from Ref.~\cite{yao2020} scaled by the magnetic Grüneisen parameter $\gamma_\mathrm{c}^\mathrm{mag}$ as described in the text. Dashed black lines mark the anomalies observed at $T_\mathrm{N}$ and $T_\mathrm{cr}$.}
    \label{fig:TE0T}
\end{figure}

The out-of-plane thermal expansion $dL_\mathrm{c}/L_\mathrm{c}$ of \ncto\ at $B = 0$~T, shown in Fig.~\ref{fig:TE0T}a, demonstrates a distinct kink at the long-range magnetic ordering temperature $T_\mathrm{N}$ as well as a broad step-like feature at lower temperature around $T_\mathrm{cr}$.~\cite{arneth2025} Consequently, the evolution of antiferromagnetic long-range order is signalled by a sharp $\lambda$-shaped anomaly in $\alpha_\mathrm{c}$ (see the inset), thereby evidencing the presence of pronounced magnetoelastic coupling. According to the Ehrenfest relation $\partial T_\mathrm{N}/\partial p_\mathrm{c} = T_\mathrm{N}V_\mathrm{m}\Delta\alpha_\mathrm{c}/\Delta C_\mathrm{p}$,~\cite{ehrenfest,moin2016}, where $V_\mathrm{m}$ denotes the molecular volume, the sign of the anomalous length changes, i.e., shrinking of the $c$ axis upon entering the antiferromagnetic state, directly implies a positive uniaxial pressure dependence of $T_\mathrm{N}$ for compression along the $c$ axis. Quantifying the size of the actual jumps in $\alpha_\mathrm{c}$ and $c_\mathrm{p}$ from Ref.~\cite{yao2020} by means of area conserving constructions~\cite{meingast1990,majumder2018,gries2022}, as depicted in the inset of Fig.~\ref{fig:TE0T}(a) yields $\partial T_\mathrm{N}/\partial p_\mathrm{c} = 0.26(5)$~K/GPa.

A more detailed investigation of the uniaxial pressure effects in \ncto\ is achieved by performing a Grüneisen analysis~\cite{barronwhite,franse1989,klingeler2006}, for which we extract the magnetic contribution to the thermal expansion coefficient using $\alpha_\mathrm{c}^\mathrm{mag} = \alpha_\mathrm{c}(B) -  \alpha_\mathrm{c}(B\parallel a^* = 15\,\mathrm{T})$ as described in Ref.~\cite{arneth2025}. This approach utilizes that magnetic contributions to the thermal expansion are strongly suppressed by $B||a^*$, such that $\alpha_\mathrm{c}(B\parallel a^* = 15\,\mathrm{T})\simeq \alpha_{\rm c}^{\rm non-mag}$, with $\alpha_{\rm c}^{\rm non-mag}$ being the non-magnetic (phononic) contributions to the thermal expansion coefficient. As visible in Fig.~\ref{fig:TE0T}b, $\alpha_\mathrm{c}^\mathrm{mag}$ and $c_\mathrm{p}^\mathrm{mag}$ exhibit a similar temperature dependence in the vicinity of the anomaly at $T_\mathrm{N}$, and can, hence, be scaled onto each other by a temperature-independent magnetic Grüneisen parameter $\gamma_\mathrm{c}^\mathrm{mag} = \alpha_\mathrm{c}^\mathrm{mag}/c_\mathrm{p}^\mathrm{mag} = 0.12(2)\,\mathrm{mol/MJ}$ around \tn . This experimental observation implies that the emergence of antiferromagnetic long-range order in \ncto\ is driven by a single dominant energy scale $\epsilon^*$. The uniaxial pressure dependence of $\epsilon^*$ is given by the Grüneisen relation $V_\mathrm{m}\gamma_\mathrm{c}^\mathrm{mag} = \partial\ln\epsilon^*/\partial p_\mathrm{c} \simeq 1\,\%/\mathrm{GPa}$. Since only one energy scale significantly contributes to the entropy changes at $T_\mathrm{N}$, it is straightforward to identify $\epsilon^*$ with the magnetic ordering temperature, i.e., $\epsilon^* \propto T_\mathrm{N}$. Doing so yields the uniaxial pressure dependence of $\partial T_\mathrm{N}/\partial p_\mathrm{c} = 0.28(5)$~K/GPa, which is in excellent agreement with the value obtained from the Ehrenfest analysis. 

In contrast to the observation of a temperature-independent magnetic Grüneisen parameter around $T_\mathrm{N}$, $\alpha_\mathrm{c}^\mathrm{mag}$ and $c_\mathrm{p}^\mathrm{mag}$ follow a significantly different temperature dependence below $T \simeq 18$~K, such that the Grüneisen scaling using only a single parameter $\gamma_\mathrm{c}^\mathrm{mag}$ is not possible at low temperatures. Hence, our data imply that the anomalous entropy changes around $T_\mathrm{cr}$ can not exclusively be attributed to the magnetic interplane interaction as discussed in Ref.~\cite{yao2020} but must arise from at least two competing energy scales.

The small positive pressure dependence of $T_\mathrm{N}$ signals that antiferromagnetic long-range order in \ncto\ is weakly stabilised upon compression of the lattice along the $c$ axis. As discussed for other 2D quantum magnets~\cite{arneth2022,spachmann2023,huang2023}, this likely results from increasing intralayer coupling, which presumably resembles the dominant energy scale $\epsilon^*$, due to changes in the Co-O-Co bond angle. An enhancement of the magnetic interplane interaction as a result of rising orbital overlap also seems plausible, but is rather unlikely to significantly contribute to $T_\mathrm{N}$.~\cite{viciu2007,bera2017} In contrast to the small positive value of $\partial T_\mathrm{N}/\partial p_\mathrm{c}$ determined for \ncto\ here, compressive uniaxial strain perpendicular to the honeycomb planes of the closely related Kitaev candidate $\alpha-$RuCl$_3$ was found to shift the balance of magnetic couplings in favour of the Kitaev exchange, and thereby suppresses antiferromagnetic order at a rate of $\partial T_\mathrm{N}/\partial p_\mathrm{c} \simeq -12\,\mathrm{K/GPa}$~\cite{kaib2021,gass2020,he2018}, which is almost two orders of magnitude larger than reported here for \ncto . This discrepancy likely results from the important role of interlayer interactions in the exchange frustration mechanism found for $\alpha-$RuCl$_3$~\cite{balz2021}, whereas magnetic coupling in \ncto\ is effectively two-dimensional. Hence, our results emphasize that, despite showing similar physical properties as $\alpha-$RuCl$_3$, uniaxial out-of-plane pressure is \textit{not} a suitable external perturbation to tune \ncto\ into a QSL state.

\begin{figure}[]
    \centering
    \includegraphics[width = 0.5\columnwidth]{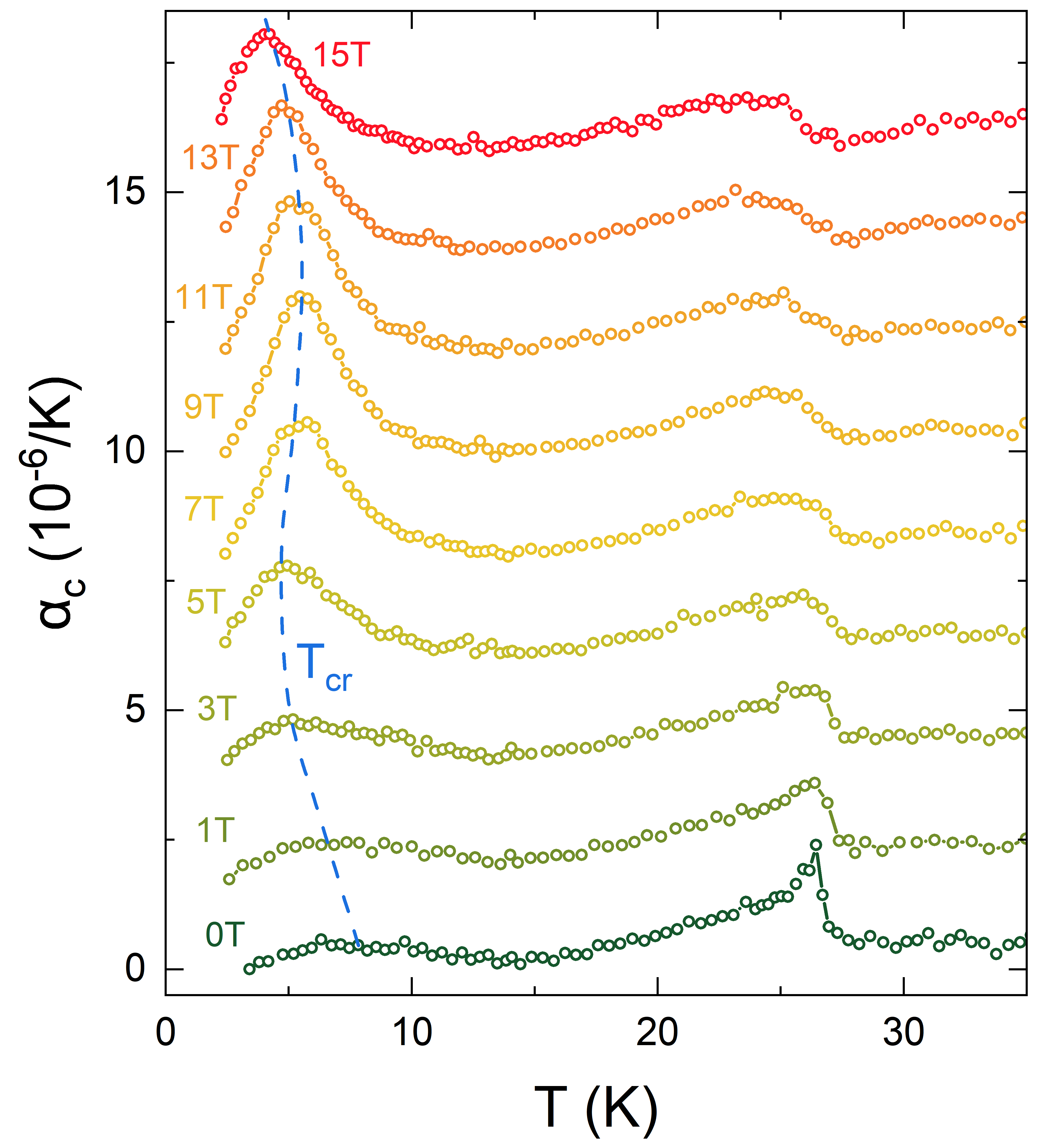}
    \caption{Out-of-plane thermal expansion coefficient at selected magnetic fields applied $B\parallel\mathrm{c}$. The data are shifted along the ordinate for better visibility. The dashed line qualitatively marks the field dependence of \tc . }
    \label{fig:TEinfield}
\end{figure}

\begin{figure}[]
    \centering
    \includegraphics[width = 0.6\columnwidth]{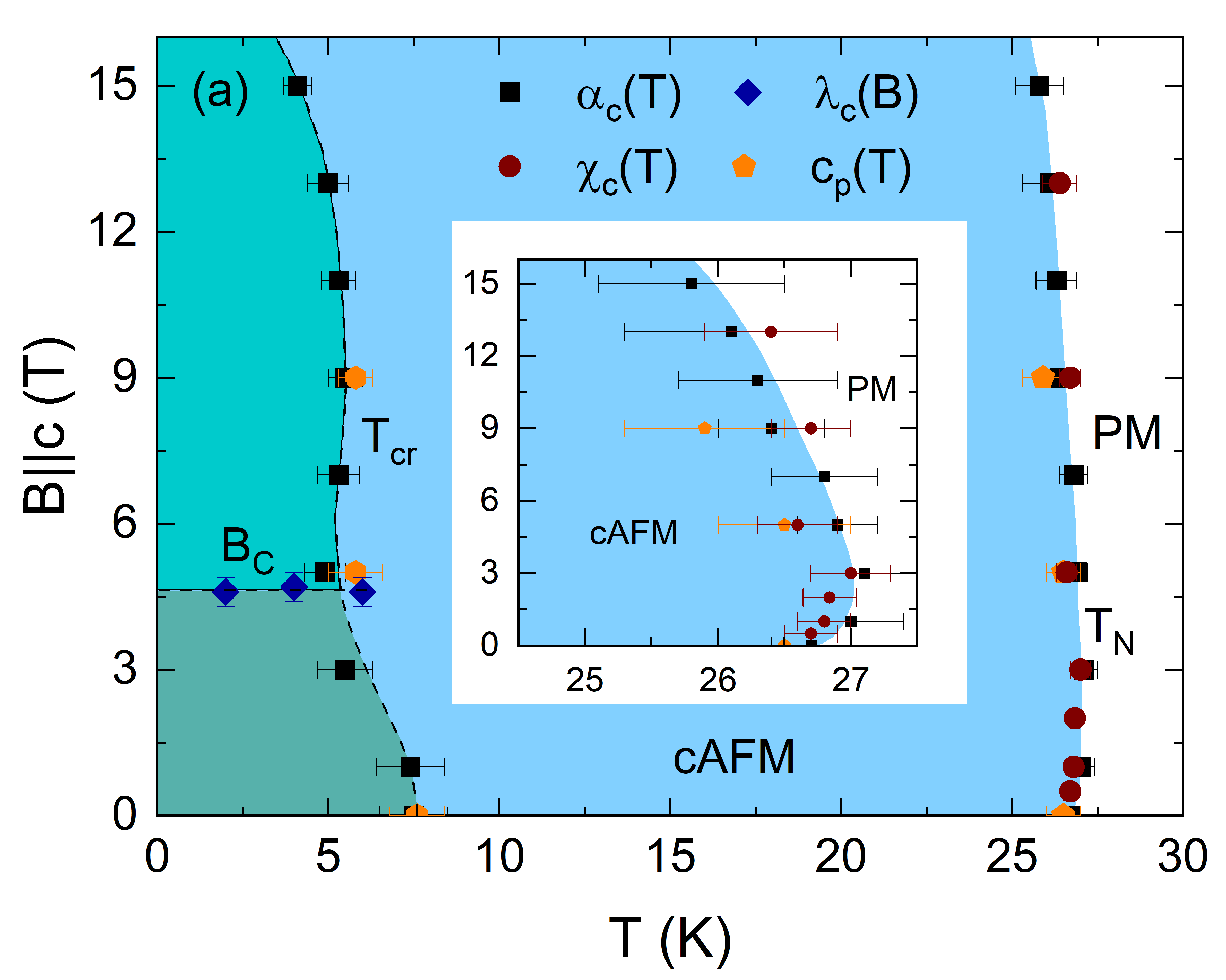}
    \caption{Magnetoelastic phase diagram of \ncto\ for $B\parallel\mathrm{c}$ as constructed from dilatometry and magnetisation data. For comparison, anomaly position in specific heat data from Ref.~\cite{yao2020} are shown, too. 
    PM and cAFM denote the paramagnetic and canted antiferromagnetic phase, respectively. All lines and coloured areas are guide to the eye.}
    \label{fig:PhD}
\end{figure}

The effects of magnetic fields applied perpendicular to the Co-layers on the thermal expansion coefficient are displayed in Fig.~\ref{fig:TEinfield}. While both anomalies at \tn\ and \tc\ are visible for all measured fields up to $15$~T and no new features occur in \ac , the data illustrate quite different field effects on the two anomalies. In agreement with reported magnetic susceptibility measurements~\cite{yao2020, zhang2024}, the thermal expansion data in Fig.~\ref{fig:TEinfield} demonstrate only a small suppression of \tn\ in an applied field $B\parallel c=15$~T. This contradicts the effect of magnetic fields applied parallel to the honeycomb planes which yield the complete suppression of 
long-range magnetic order already for $B\perp c\simeq 10$~T. Our data hence confirm a strongly anisotropic magnetic field effect on the ground state of \ncto . At the same time, the data in Fig.~\ref{fig:TEinfield} show that the pronounced $\lambda$-shaped anomaly associated with the evolution of long-range antiferromagnetic order evolves into a broader step-like feature indicating the suppression of critical fluctuations. In contrast to the only small changes at \tn , the broad hump centred around \tc\ is strongly affected by magnetic fields along the $c$ axis. Specifically, up to $B = 9$~T, the feature becomes significantly more pronounced and turns into a sharp, almost symmetric peak which again starts to decrease for even higher fields. Simultaneously, \tc\ follows a non-monotonous magnetic field dependence, that is, the associated hump shifts to lower temperatures in both small and high magnetic fields, whereas a positive $\partial T_\mathrm{cr}/\partial B$ is observed in the field range $3\,\mathrm{T} \lesssim B \lesssim 9\,\mathrm{T}$ (see the dashed line in Fig.~\ref{fig:TEinfield} and Fig.~\ref{fig:PhD}).

The quantitatively and qualitatively contrasting field dependencies of both the size and the positions of the anomalies at \tn\ and \tc\ clearly imply their different microscopic origin and, hence, the presence of multiple competing energy scales. In particular, our data evidence that the competing energy scales are differently affected by magnetic fields $B||c$. While the ratio of the jumps $\Delta \alpha_\mathrm{c}$ and $\Delta c_\mathrm{p}$ from Ref.~\cite{yao2020} at \tn\ is roughly constant in the different magnetic fields and, thereby, indicates a rather field-independent pressure dependence $\partial T_\mathrm{N}/\partial p_\mathrm{c}$, the findings for \tc\ are very different: Exploiting the Grüneisen relation implies the increase of $\partial \ln T_\mathrm{cr}/\partial p_\mathrm{c}$ with the external magnetic field by a factor of 3 in $B=5$~T and of nearly 5 in $B=9$~T (see Tab.~\ref{tab:Pdep}). The strong variation of the uniaxial pressure coefficient of \tc\ upon application of magnetic fields $B||c$ can be directly read-off the data in Fig.~\ref{fig:TEinfield} where the associated peak in \ac\ is roughly six times larger at $9$~T than in zero field while the corresponding feature in $c_\mathrm{p}$ only doubles in size~\cite{yao2020}. The increasing pressure coefficient implies that the competing energy scale evolving around \tc\ becomes more susceptible to out-of-plane lattice strain when magnetic fields $B\parallel c$ are applied which qualitatively corroborates its assignment with short-ranged out-of-plane correlations.

\begin{table}[h]
    \centering
    \begin{tabular}{c|c|c}
        $B \parallel c$ & $\partial T_\mathbf{N}/\partial p_\mathrm{c}$ (K/GPa) & $\partial \ln T_\mathrm{cr}/\partial p_\mathrm{c}$ ($\%$/GPa) \\ \hline
        0 & 0.28(5) & 6.1(17) \\
        5 & 0.27(5) & 17(5) \\
        9 & 0.28(6) & 27(6) \\
    \end{tabular}
    \caption{Pressure dependecies of \tn\ and \tc\ at different magnetic fields applied $B \parallel c$ (see the text).}
    \label{tab:Pdep}
\end{table}

\begin{figure}[]
    \centering
    \includegraphics[width = 0.5\columnwidth]{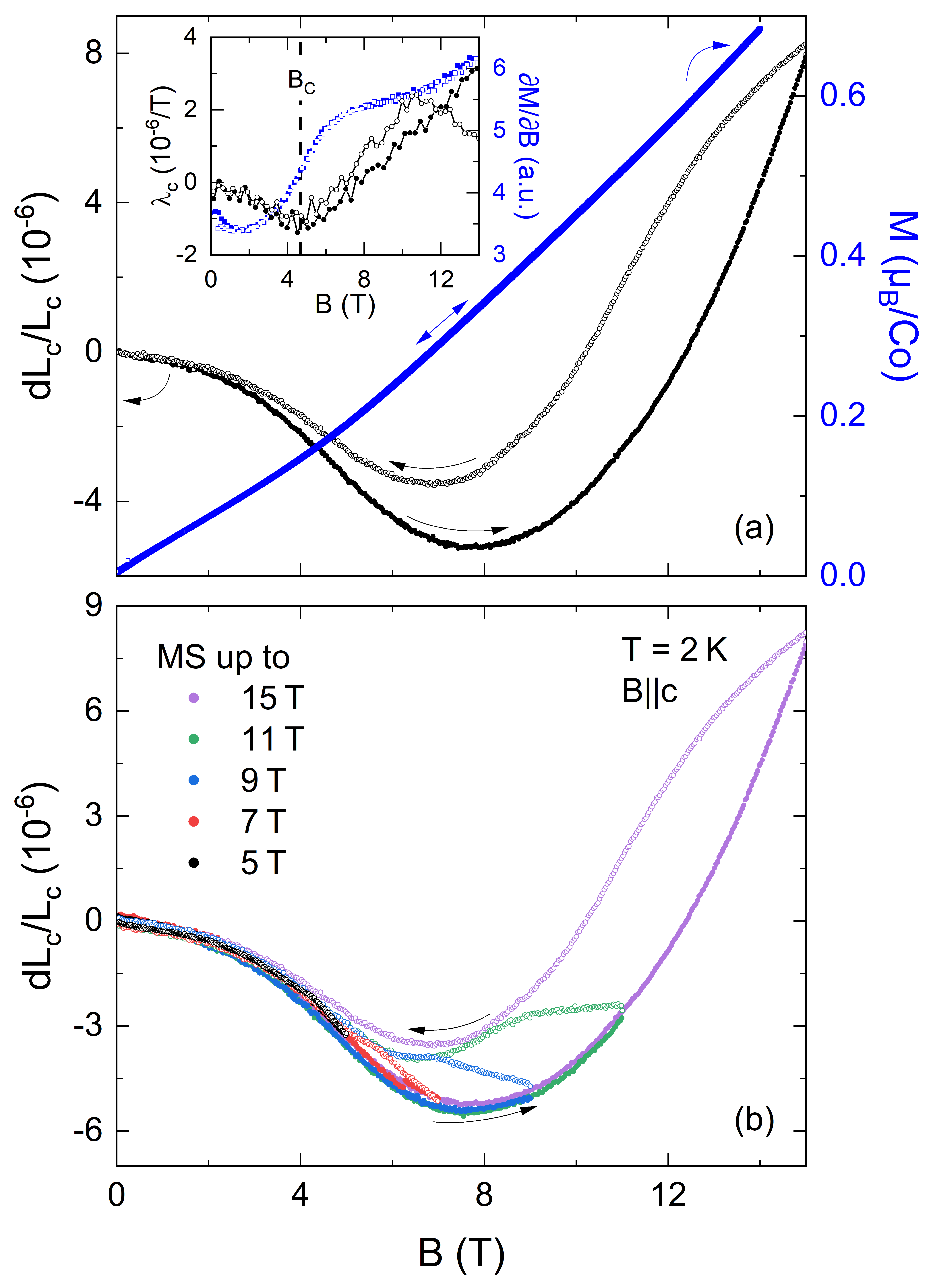}
    \caption{(a) Out-of-plane magnetostriction and isothermal magnetisation for $B\parallel\mathrm{c}$ at $T = 2$~K. The inset displays the corresponding magnetostriction coefficient $\lambda_\mathrm{c}$ and susceptibility $\partial M/\partial B$. (b) Out-of-plane magnetostriction at $T = 2$~K up to different maximum magnetic fields $B \parallel c$. Closed (open) symbols depict up-(down-)sweeps of the magnetic field.}
    \label{fig:MSMB}
\end{figure}

Magnetoelastic coupling in \ncto\ is further investigated by our magnetostriction studies. The field-induced length changes up to $14$~T measured at $T = 2$~K are shown in Fig.~\ref{fig:MSMB}a. For comparison, the isothermal magnetisation obtained at the same temperature is shown in the diagram as well. We observe the following main features: Starting from zero field, $dL_\mathrm{c}/L_\mathrm{c}(B)$ is concave and continuously decreases with magnetic field. At $B_\mathrm{C} \simeq 4.6$~T, magnetostriction changes its curvature from concave to convex but continues to decrease further until a minimum appears around $B_{\rm min}\simeq 8$~T. Above 8~T, the $c$ axis continuously expands up to the highest measured field. The inflection point in $dL_\mathrm{c}/L_\mathrm{c}(B)$ coincides with a kink in the isothermal magnetisation at $B_{\rm C}$, which has been observed but is only scarcely discussed in the literature~\cite{zhang2024,zhou2024_arxiv}. The correspondence of lattice and magnetization changes is better visible in the respective first derivatives, i.e., in the magnetostriction coefficient $\lambda_\mathrm{c}$ and in the differential magnetic susceptibility $\partial M/\partial B$ shown in the inset of Fig.~\ref{fig:MSMB}, where the minimum in $\lambda_\mathrm{c}$ coincides with a distinct jump in $\partial M/\partial B$. Simultaneously, a low-temperature upturn of the static magnetic susceptibility $\chi_\mathrm{c}(T)=M_{\rm c}(T)/B$ is found only for $B \gtrsim 5$~T (see Fig.~\ref{fig:highBChi_supp}), which has been heuristically related to a change of the ferrimagnetic spin canting along $c$~\cite{yao2020}. In agreement with recent electron spin resonance spectroscopy studies revealing the softening of a magnon excitation mode and, hence, symmetry-breaking in a similar field region~\cite{bera2023,bischoff2025}, the here presented magnetic data exhibit characteristics of a continuous (second-order) phase transition at \bc . On the other hand, the associated magnetostrictive response implies a weak discontinuous character at most, which renders the assignment of the observed anomalies to a second-order thermodynamic phase boundary unclear. The position of \bc\ is virtually temperature-independent up to $6$~K, i.e., around \tc ($B\simeq 5$~T), above which no associated feature can be discerned (see Fig.~\ref{fig:MShighT_supp}). Following the interpretation of \bc\ marking a phase transition, the absence of corresponding features at temperatures above \tc\ indicates that the latter signals a phase boundary as well (cf.~the magnetoelastic phase diagram in Fig.~\ref{fig:PhD}).

In contrast to the results of pulsed magnetisation measurements in Ref.~\cite{zhang2024}, we neither observe a peak centred around $9$~T in our quasi-static $\partial M/\partial B$ data nor identify any corresponding features in the lattice parameters. We conclude that the reported characteristics may be associated with dynamic effects at higher sweep rates. Notably, our $c$ axis magnetostriction data in Fig.~\ref{fig:MSMB} display pronounced hysteresis from $B \simeq 2$~T up to the highest field of $14$~T while the up- and down-sweep of the isothermal magnetisation perfectly coincide within the resolution of our experiment. A similar finding, i.e., hysteretic behaviour in the field-induced length changes, but not in the magnetic properties, was discussed to arise from the formation of structural domains also for fields along the $a^*$ axis~\cite{arneth2025}. Our data, hence, imply that structural domains are present for $B\parallel c$ as well. Interestingly, hysteresis in $dL_\mathrm{c}/L_\mathrm{c}(B)$ only occurs when the magnetic field is increased beyond \bc\ (see Fig.~\ref{fig:MSMB}b). Following the putative assignment of \bc\ to a phase boundary, this observation would imply that structural domains appear in the presumed high-field phase. Such domain effects could, therefore, explain the observed unusual shape of the anomaly marking \bc\ in the magnetostriction. Note here that, while the minimum in $dL_\mathrm{c}/L_\mathrm{c}$ shifts by roughly $1$~T between up- and down-sweep, the position of \bc\ is independent of the direction of the magnetic field sweep.

The anomaly positions extracted from our magnetic and dilatometric data are summarised in the magnetoelastic phase diagram shown in Fig.~\ref{fig:PhD}. For reference, we also include the results of the specific heat capacity measurements from Ref.~\cite{yao2020}. As discussed earlier, the phase boundary marked by \tn\ separating the paramagnetic (PM) from the canted antiferromagnetic (cAFM) state is rather robust against magnetic fields applied along the $c$ axis. In numbers, an external field of $15$~T reduces the N\'{e}el temperature by only about $1$~K. A more detailed investigation, moreover, reveals that \tn\ does not decrease continuously but two different regimes can be identified (see the inset of Fig.~\ref{fig:PhD}). At high magnetic fields $B \gtrsim 5$~T, the magnetic order is suppressed by magnetic fields as expected for long-range antiferromagnetic order. In contrast, $\partial T_\mathrm{N}/\partial B$ is positive for small fields up to at least $3$~T which likely arises due to ferrimagnetic spin canting along $c$ in the magnetic ground state. Note that the observed positive field dependence of \tn\ is thermodynamically linked to the sign of the corresponding anomalies in the magnetic susceptibility by means of the Ehrenfest relation $\partial T/\partial B = -TV\Delta(\partial M/\partial T)/\Delta c_\mathrm{p}$~\cite{ehrenfest}. The superposition of these two regimes leads to a maximum in \tn\ at $B\simeq 4$~T. At this point, it is noteworthy that the sign change of $\partial T_\mathrm{N}/\partial B$ occurs at fields only slightly smaller than \bc . Thus, one might speculate that the low-temperature features at \bc\ and the changes in the slope \tc ($B$) in this field range result from the same microscopic mechanism (possibly a change of spin canting along $c$) as the sign change in $\partial T_\mathrm{N}/\partial B$.

The magnetic field dependence of \tc\ is also not monotonous but can be separated into three regions. For $B \lesssim 3-5$~T, the observed peaks in $\alpha_\mathrm{c}$ and $c_\mathrm{p}$ shifts to smaller temperatures with increasing external field. Subsequently, \tc\ increases (this behaviour is better visible directly in the $\alpha_\mathrm{c}$ data in Fig.~\ref{fig:TEinfield}) at intermediate fields up to $9$~T until being suppressed again by even higher magnetic fields. The first crossover region also falls into the field range of \bc\ hinting at an intertwining of the two phenomena. The second sign change of $\partial T_\mathrm{cr}/\partial B$ is not accompanied by any distinct features in the here presented static thermodynamic properties but might be related to the peak occurring around $9$~T in the pulsed magnetisation measurements, which would imply a spin dynamic origin~\cite{zhang2024}. The observation of clear anomalies in the thermodynamic response functions strongly implies that \tc\ and \bc\ are associated with phase transitions, the exact nature of which remains an open question.

\section{Conclusions}

In conclusion, we report a detailed dilatometric study on the Kitaev spin liquid candidate \ncto\ up to high magnetic fields applied $B \parallel c$. In contrast to the findings for the closely related $\alpha$-RuCl$_3$, uniaxial compression perpendicular to the honeycomb planes is demonstrated to stabilise the antiferromagnetic ground state in \ncto . Moreover, the calculated pressure coefficient $\partial T_\mathrm{N}/\partial p_\mathrm{c} = 0.28(5)$~K/GPa is almost two orders of magnitude smaller than in $\alpha$-RuCl$_3$ which renders tensile out-of-plane strain a rather ineffective perturbation to drive \ncto\ into a spin disordered phase. While the entropy changes around \tn\ are dominated by a single energy scale $\epsilon^*$, the failure of  Grüneisen scaling around \tc\ implies competing interactions at lower temperatures. Magnetic fields applied parallel to the honeycomb layers do not significantly suppress the magnetic order, but the low-temperature energy scale is found to become increasingly sensitive to lattice changes at high fields. The conclusion of strong coupling between spin and lattice degrees of freedom is further corroborated by our magnetostriction measurements showing that both the isothermal magnetization and the field-induced length changes exhibit a clear feature at \bc\ which putatively arises from a change in the spin canting along the $c$ axis. The magnetoelastic phase diagram reveals sign changes in the field dependencies of both \tn\ and \tc\ connected to \bc . Summarising, our investigations provide a deeper understanding of the role of magnetoelastic coupling in the ground state properties of \ncto .


\begin{acknowledgments}
We acknowledge support by Deutsche Forschungsgemeinschaft (DFG) under Germany’s Excellence Strategy EXC2181/1-390900948 (the Heidelberg STRUCTURES Excellence Cluster). JA acknowledges support by the IMPRS-QD Heidelberg. KYC was supported by the National Research Foundation (NRF) of Korea (Grant Nos.~2020R1A5A1016518 and RS-2023-00209121). R.S.~acknowledges the financial support provided by the Ministry of Science and Technology in Taiwan under project numbers NSTC 111-2124-M-001-007, Financial support from the Center of Atomic Initiative for New
Materials (AI-Mat), (Project No.~108L9008) and Academia Sinica for the budget of AS-iMATE-113-12.
\end{acknowledgments}


\hfill

\appendix
\section*{Appendix}\label{Sec_Appendix}

\renewcommand{\thefigure}{A\arabic{figure}}
\setcounter{figure}{0}  

\section{Static magnetic susceptibility}\label{sec:a1}

\begin{figure}[h]
    \centering
    \includegraphics[width = 0.5\columnwidth]{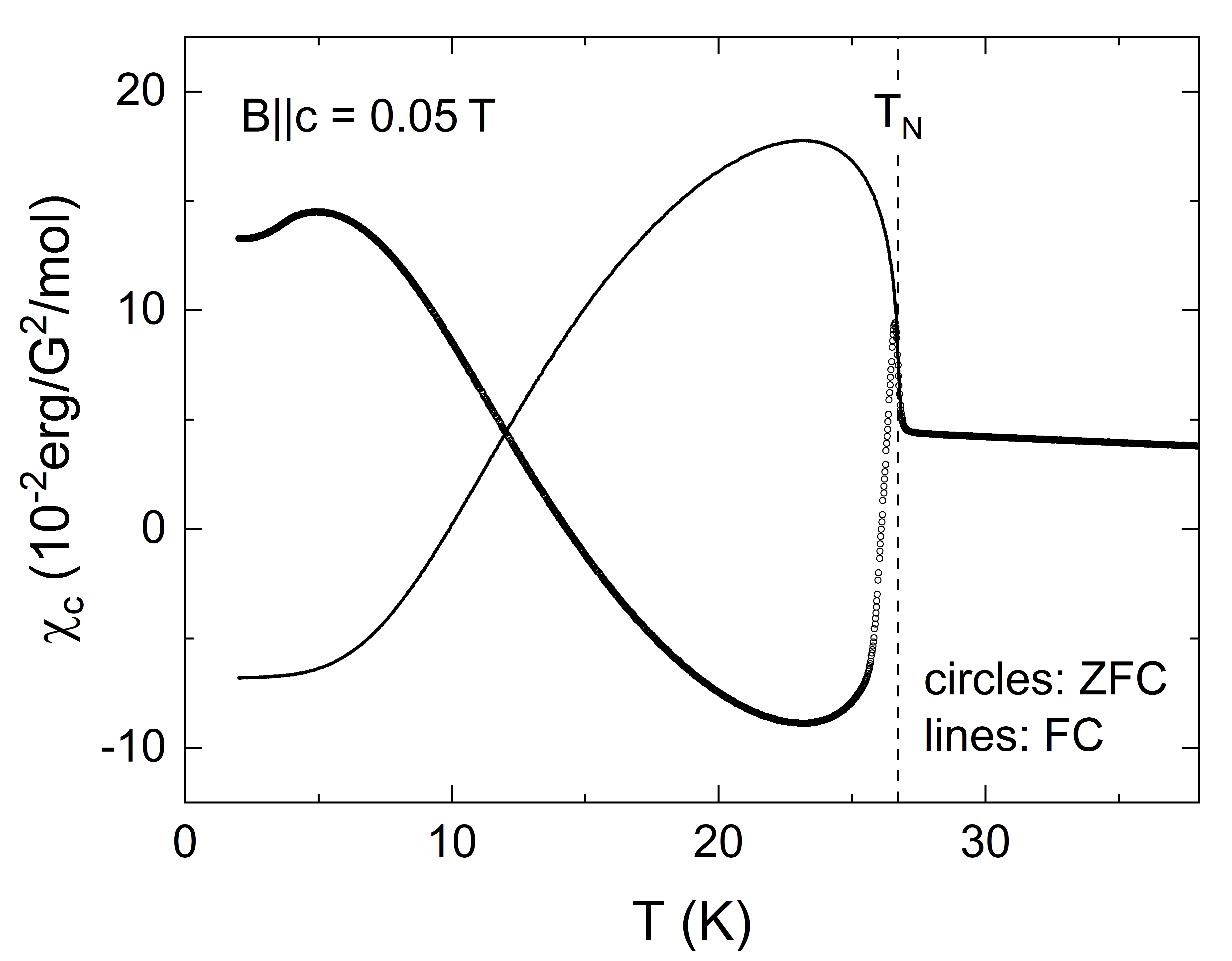}
    \caption{Zero field cooled (ZFC) and field cooled (FC) out-of-plane magnetic susceptibility at $B = 0.05$~T. The dashed vertical line marks the emergence of magnetic order at \tn .}
    \label{fig:lowBChi_supp}
\end{figure}

The out-of-plane static magnetic susceptibility of \ncto\ at $B = 0.05$~T, shown in Fig.~\ref{fig:lowBChi_supp}, is characterised by a pronounced kink at $T_\mathrm{N} = 26.8$~K marking the evolution of canted antiferromagnetic long-range order. Due to unequal spin canting on the two different Co-sites, $\chi_c$ exhibits a rather ferrimagnetic behaviour, i.e., strong hysteresis below \tn\ and initially negative net magnetisation after field cooling. In addition to the prominent anomaly at \tn , a small jump can be identified at low temperatures in the ZFC data. The absence of this feature in the FC data implies that it is associated with dynamic effects, such as domain wall movement. Importantly, we do not observe any additional anomalies in the susceptibility that might correspond to putative impurity phases, such as the antiferromagnetic polymorph reported in Ref.~\cite{dufault2023}.

\begin{figure}[]
    \centering
    \includegraphics[width = 0.5\columnwidth]{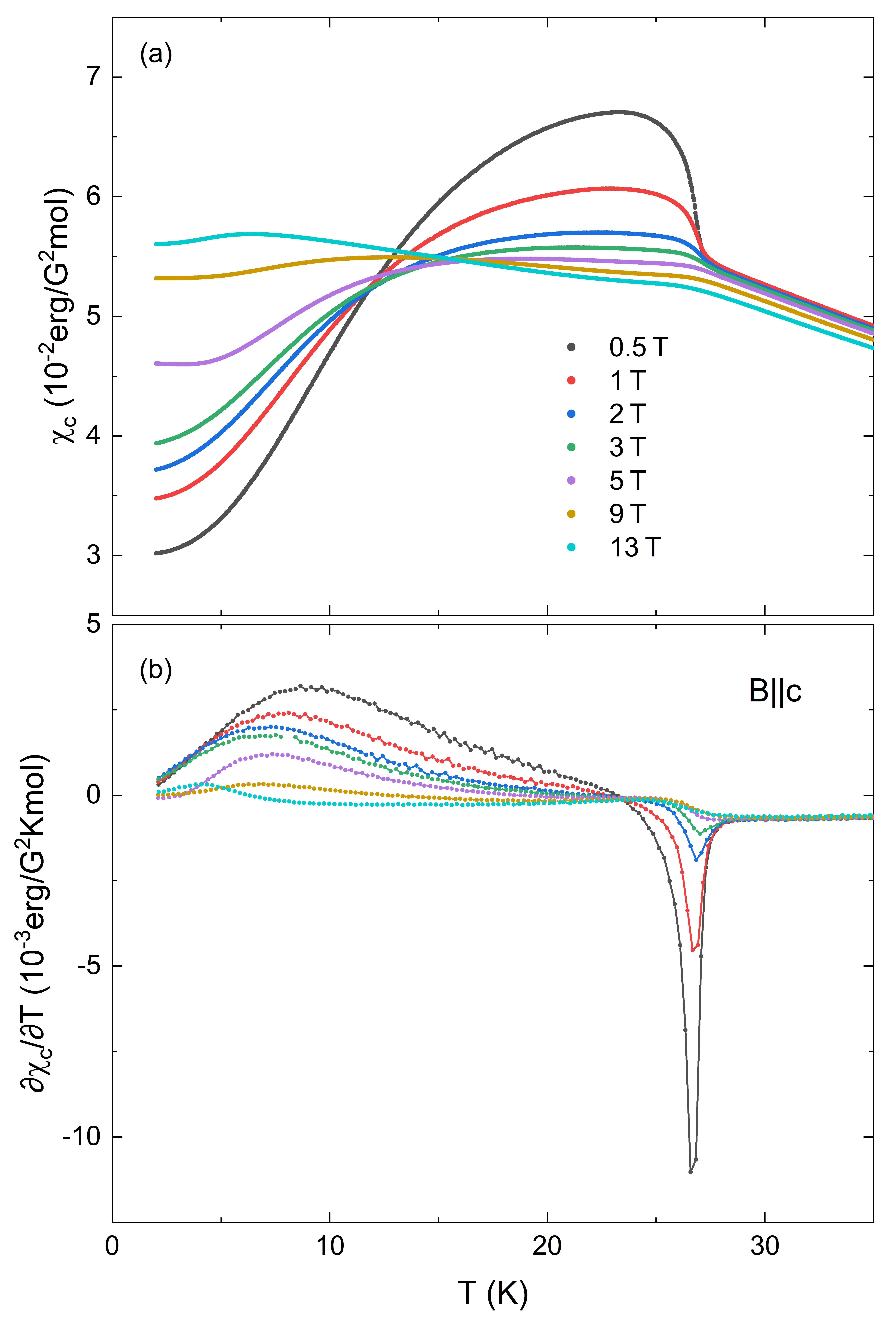}
    \caption{Out-of-plane magnetic susceptibility (a) and its temperature derivative (b) at various selected external magnetic fields. All measurements were performed following a field-cooled protocol.}
    \label{fig:highBChi_supp}
\end{figure}

The effects of higher magnetic fields $B \parallel c$ on $\chi_c$ are depicted in Fig.~\ref{fig:highBChi_supp}. In agreement with the thermal expansion data, \tn\ is only hardly affected by the external fields. However, the corresponding anomaly is found to exhibit a different shape in small and high fields: While \tn\ is marked by a pronounced upwards kink, i.e., $\Delta\partial M/\partial B > 0$ upon heating, for $B \lesssim 3$~T, the anomaly exhibits the opposite sign for $B \gtrsim 5$~t. Utilizing the Ehrenfest relation, this straightforwardly corroborates the findings of the magnetoelastic phase diagram as described in the main text. Simultaneously, $\chi_c$ exhibits a small upturn at low temperatures for $B \geq 5$~T, but not for $B \leq 3$~T, which might be related to the critical field \bc\ and is discussed in more detail in the main text.

\section{Magnetostriction at higher temperatures}\label{sec:a2}

\begin{figure}[]
    \centering
    \includegraphics[width = 0.5\columnwidth]{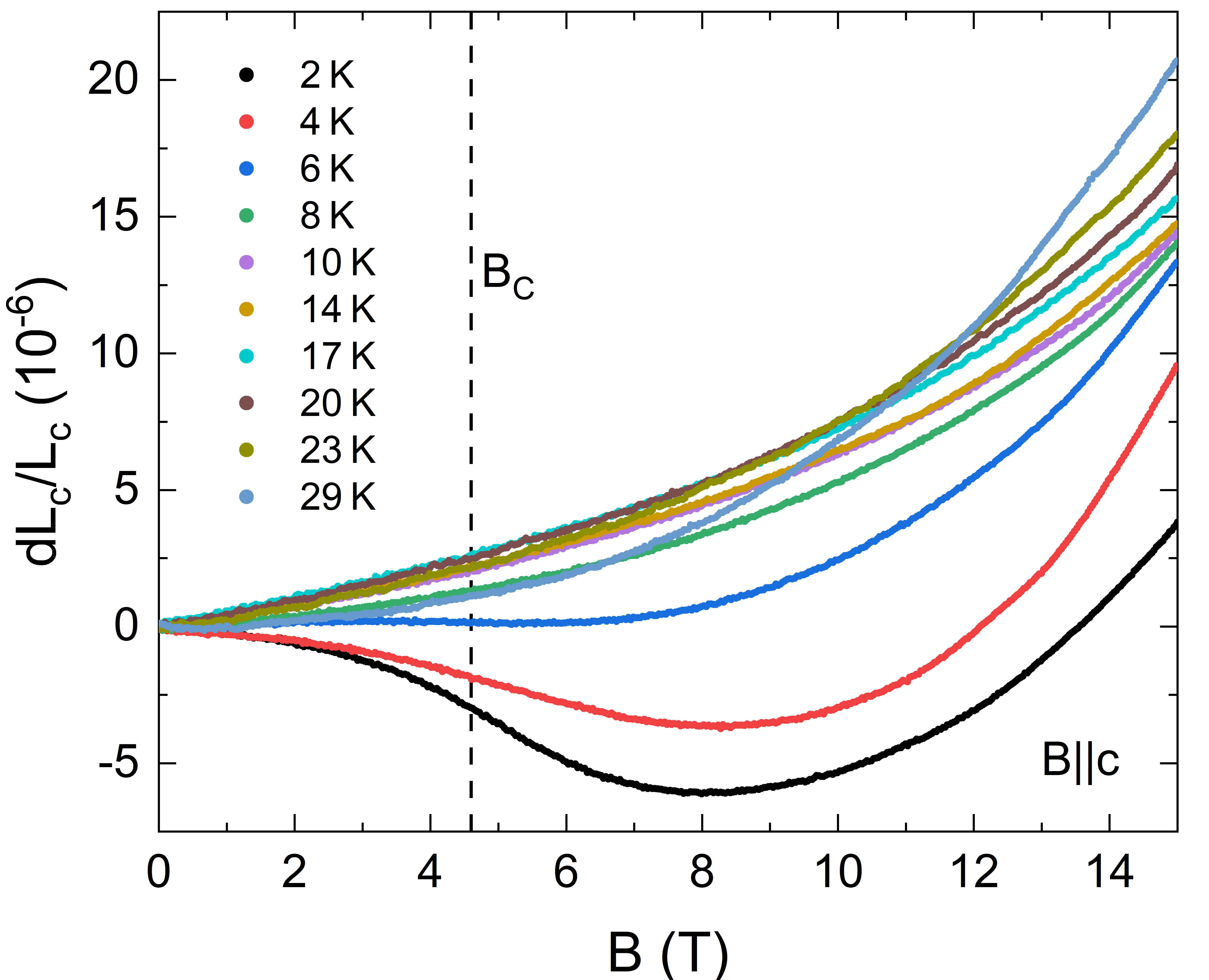}
    \caption{Out-of-plane magnetostriction for $B \parallel c$ measured at selected temperatures. For better visibility only the up-sweeps are shown. The black dashed line marks the critical field \bc .}
    \label{fig:MShighT_supp}
\end{figure}

Fig.~\ref{fig:MShighT_supp} shows the out-of-plane magnetostriction of \ncto\ at selected temperatures measured for increasing magnetic fields $B \parallel c$. On increasing the temperature, shrinking of the $c$ axis at small fields becomes less pronounced and both the minimum and the inflection point in $dL_\mathrm{c}/L_\mathrm{c}$ gradually broaden until they are no longer discernable for $T \gtrsim 8$~K. At temperatures higher than $6$~K, magnetostriction is positive in the entire measured field range. While the position of $B_\mathrm{min}$ shifts with varying temperature, \bc\ is essentially temperature independent.

\bibliography{litNCTO}

\end{document}